

\input phyzzx
\input epsf
%
\catcode`@=11
%
%
\newtoks\KUNS
\newtoks\HETH
\newtoks\monthyear
\newtoks\HEPN
\Pubnum={KUNS~\the\KUNS\cr HE(TH)~\the\HETH\cr \the\HEPN}
\monthyear={\monthname,\ \number\year}
\def\p@bblock{\begingroup \tabskip=\hsize minus \hsize
   \baselineskip=1.5\ht\strutbox \topspace-2\baselineskip
   \halign to\hsize{\strut ##\hfil\tabskip=0pt\crcr
   \the\Pubnum\cr \the\monthyear\cr }\endgroup}
\def\bftitlestyle#1{\par\begingroup \titleparagraphs
     \iftwelv@\fourteenpoint\else\twelvepoint\fi
   \noindent {\bf #1}\par\endgroup }
\def\title#1{\vskip\frontpageskip \bftitlestyle{#1} \vskip\headskip }
%
%
\def\Kyoto{\address{Department of Physics,~Kyoto University \break
                            Kyoto~606,~JAPAN}}

%
%
\paperfootline={\hss\iffrontpage\else\ifp@genum%
                \tenrm --\thinspace\folio\thinspace --\hss\fi\fi}
\footline=\paperfootline
%
%

%
\def\journal#1&#2(#3){\begingroup \let\journal=\dummyj@urnal
    \unskip, \sl #1\unskip~\bf\ignorespaces #2\rm
    (\afterassignment\j@ur \count255=#3) \endgroup\ignorespaces }
\def\andjournal#1&#2(#3){\begingroup \let\journal=\dummyj@urnal
    \sl #1\unskip~\bf\ignorespaces #2\rm
    (\afterassignment\j@ur \count255=#3) \endgroup\ignorespaces }
\def\andvol&#1(#2){\begingroup \let\journal=\dummyj@urnal
    \bf\ignorespaces #1\rm
    (\afterassignment\j@ur \count255=#2) \endgroup\ignorespaces }
\def\NP{Nucl.~Phys. }
\def\PR{Phys.~Rev. }

\def\PL{Phys.~Lett. }
\def\PTP{Prog.~Theor.~Phys. }

\def\MPL{Mod.~Phys.~Lett.}

%
%
\def\acknowledge{\par\penalty-100\medskip \spacecheck\sectionminspace
   \line{\hfil ACKNOWLEDGEMENTS\hfil}\nobreak\vskip\headskip }

\KUNS={1326}     
\HETH={95/05}   
\HEPN={hep-ph/9503274}

\def\uy{U(1)_Y}
\def\ua{U(1)}
\def\ub{SU(2)}
\def\uc{SU(3)}
\def\Tt{{\tilde T}}
\def\Ht{{\tilde H}}
\def\dg#1{#1^\circ}

\REF\ETC{
S.~Dimopplous and L.~Susskind \journal \NP &B155 (79) 237. \nextline
E.~Eichten and K.~Lane \journal \PL &90B (80) 125.}

\REF\TopCon{
V.~A.~Miransky, M.~Tanabashi and K.~Yamawaki
  \journal \PL &B221 (89) 177;
  \andjournal Mod. Phys. Lett. &A4 (89) 1043.\nextline
Y.~Nambu, EFI preprint 89-08.\nextline
W.~A.~Bardeen, C.~T.~Hill and M.~Lindner
  \journal \PR &D41 (90) 1647.}

\REF\NNT{
Y.~Nagoshi, K.~Nakanishi and S.~Tanaka \journal \PTP &85 (91) 131.}

\REF\NT{
K.~Nakanishi and S.~Tanaka \journal \PTP &87 (92) 479.}

\REF\NNTb{
Y.~Nagoshi, K.~Nakanishi and S.~Tanaka \journal \PTP &90 (93) 1311.}

\REF\BKM{
M.~Bando, T.~Kugo and G.M.~Mitchard, Int. Workshop on Electroweak
Symmetry Breaking (World Scientific, 1992), p.51.}

\REF\Yama{
K.~Yamawaki, Proc. 1990 Int. Workshop on Strong Coupled Gauge
Theories and Beyond, Nagoya, July 28-31, 1990, p.13.}

\REF\BKS{
M.~Bando, T.~Kugo and K.~Suehiro \journal \PTP &85 (91) 1299.}

\REF\KSY{
K-I.~Kondo, S.~Shuto and K~Yamawaki \journal \MPL &A6 (91) 3385.}

\titlepage

\title{
Dynamical Generation of CKM Mixings by Broken Horizontal Gauge
Interactions}

\author{
Yasuhiko
NAGOSHI\foot{e-mail address: nagoshi@gauge.scphys.kyoto-u.ac.jp}
and Ken
NAKANISHI\foot{e-mail address: nakanisi@gauge.scphys.kyoto-u.ac.jp}}

\Kyoto  


\abstract{
The fermion mass matrices are calculated in the framework of the
dynamical mass generation by the broken horizontal gauge interactions.
The non-proportional mass spectra between up- and down-sectors and CKM
mixings are obtained solely by radiative corrections due to the
ordinary gauge interactions.}

\endpage

\chapter{Introduction}
The standard model offers a remarkably successful description
of the gauge interactions of the particles thus far observed and
accounts extremely well for the vast amounts of high-energy particle
experimental data.
Nevertheless, it does not present any satisfactory understanding of
matter parts, involving too many arbitrary parameters,
particularly, in Higgs and Yukawa sectors.
This means that the standard model itself has no answer
for the origin of quark-lepton masses,
Cabbibo-Kobayashi-Maskawa (CKM) mixings
and the number of generations.
These fermion mass problems have been studied by many people
with various ideas.
Main purpose of these works is
to elucidate two types of hierarchies of fermion masses,
one of which is among generations and the other is among sectors
(up, down, neutrino, and electron).

One of the most attractive scenarios is a dynamical mass generation,
for example, by extended techni-color\refmark{\ETC} or top condensate
model.\refmark{\TopCon}
These models, however, do not explain the above hierarchy problems
well in spite of their successes in symmetry breaking.
We have so far been studying this problem with broken horizontal gauge
symmetry,\refmark{\NNT,\NT,\NNTb}
which is some extension of the top condensate model.
In the previous papers, it was shown that the hypercharge gauge
interaction $U(1)_Y$ plays an important role in generating hierarchy
between up- and down-sectors naturally.
The other hierarchy among generations is explained by a suitable breaking
pattern of horizontal gauge symmetry, which is, however, given by hand.
One of the purposes of the present paper is to find out the underlying
structure behind our model by studying the relation between the
breaking pattern and induced fermion mass spectra.

It was pointed out that the hypercharge interaction
does not well generate flavour mixings;\refmark{\BKM} to be precise,
the mass matrix
of up-sector $M_U$ is almost proportional to that of down-sector $M_D$,
since the effects of $U(1)_Y$ is so small.
Here, we investigate whether CKM mixings can occur or not in the above broken
horizontal gauge model.
The main task in the present paper is actually to show the breaking of
the proportionality
between $M_U$ and $M_D$ and to find the breaking pattern that causes
CKM mixings by the $U(1)_Y$ radiative corrections.
The plan of this paper is as follows.
In sect.2 we review the previous papers in brief and present a model.
In sect.3 we study eigenvalue problem, which is equivalent
to solving the mass gap equations approximately.
In sect.4 it is shown that nonlinear terms of the gap equations are
essential for CKM mixings.
In sect.5 a down-quark diagonalizing base is introduced.
In sect.6 we rewrite the down-sector gap equation as an eigenequation.
In sect.7 we show that CKM mixings can actually occur in particular cases.
Some conclusions are given in sect.8.

\chapter{Gap Equation for Fermions}
In this section, we shall review the mass gap equations for quarks
and leptons induced by the horizontal gauge interactions,
and investigate the general aspects of gap equations.

We introduce the horizontal gauge interactions
$$
L_{\rm int}=f{\bar \psi }\gamma ^\mu H^\kappa _\mu T_\kappa \psi ,
\eqn\BA
$$
in addition to the standard gauge interactions
($SU(3)_{C}\times SU(2)_L\times U(1)_Y$), where $T_\kappa $'s denote generators
of
horizontal gauge symmetry, say $SU(N)$, over $N$ generations
of fermions $\psi $.

It is assumed that the horizontal gauge symmetry breaks at the energy
scale $\Lambda $ with keeping the ordinary gauge symmetries, and the gauge
fields $H^\kappa _\mu $'s acquire a real symmetric squared-mass matrix
$\mu ^2_{\kappa \kappa '}$. Considering $SO(N^2-1)$ transformation $O_{\kappa
\kappa '}$
which diagonalizes the $\mu ^2_{\kappa \kappa '}$ with mass eigenvalues
$M^2_\kappa $,
the gauge interaction \BA\ is rewritten in terms of mass eigenmodes as
$$
f{\bar \psi }\gamma ^\mu {\tilde H}^\kappa _\mu {\tilde T}_\kappa \psi ,
\eqn\BB
$$
where
$$
{\tilde T}_\kappa =O_{\kappa \kappa '}T_{\kappa '},
\eqn\BC
$$
and
$$
{\tilde H}^\kappa _\mu =O_{\kappa \kappa '}H^{\kappa '}_\mu .
\eqn\BD
$$
The ${\tilde H}^\kappa _\mu $ denotes the mass eigen field of the horizontal
gauge boson with mass $M_\kappa $.

Before discussing the gap equation, we briefly comment on the number of
generations $N$ and the horizontal symmetry breaking. As was pointed out
in the original works of the top condensate scenario, the top quark must
be much heavier than we expect in experiments in order to supply
sufficient masses for the weak bosons. This problem can be avoided by
introducing higher generations, which may be the dominant sources of the
weak boson masses. This is the case for our model, which implies that we
must consider $N\geq 4$ models. It is assumed, however, that the higher
generations are nearly decoupled from the ordinary three generations, so
that we take the $N=3$ model and $SU(3)$ horizontal symmetry, hereafter.

The second question is what type of breaking of horizontal symmetry
should be considered.
Here, we comment only general features of the breaking
pattern. Consider a case that all horizontal gauge bosons have the same
mass. It is easily seen that the fermion mass matrix is proportional
to a unit matrix because the fermions have a global horizontal $SU(3)$
symmetry. This suggests that we should consider some hierarchical
structures of horizontal symmetry breaking for obtaining the realistic
fermion mass matrices. We hope that some group theoretical structures
underlie the hierarchical horizontal symmetry breaking, for example,
a so-called sequential breaking.

Now, let us consider the mass gap equations for fermions. Since it is
difficult to solve a Schwinger-Dyson(SD) equation, which has momentum
dependent solutions in general, especially for the present broken gauge
interactions, we adopt the following two approximations. One is
replacement of the intermediate horizontal gauge interactions by
four-fermi ones. The other is introduction of some weights into the mass gap
equations, which represent the effects of gauge boson mass. Noting
that our main purpose is investigation of texture of the fermion mass
matrices, these approximations do not influence our results.

Let us start with a simple case that all horizontal gauge bosons have the
same mass $\Lambda $. The intermediate horizontal gauge interactions are
replaced by the following four-fermi interactions,
$$
L_{\rm int}
=-{f^2\over2\Lambda ^2}({\bar \psi }\gamma _\mu T_\kappa \psi)({\bar \psi
}\gamma ^\mu T_\kappa \psi).
\eqn\BE
$$
\FIG\figA{A self energy diagram.}
The mass gap equation indicated by the diagram in Fig.\figA\ is
$$
M={f^2\over4\pi ^2}\sum_\kappa T_\kappa \left[1+{M^2\over\Lambda
^2}\ln{M^2\over\Lambda ^2}\right]M T_\kappa ,
\eqn\BF
$$
where we take horizontal breaking scale $\Lambda $ for a cut-off.
Note that $T_\kappa $'s are already represented on the gauge boson mass
diagonal base in this case.

The next step is to consider a more complicated case that the horizontal
gauge bosons have different masses. Supposing that one of horizontal
gauge bosons has infinite mass, its contribution to
the gap equation drops out. This extreme case tells that the
contribution of a heavy gauge boson to the gap equation is small, and
that of light one is large. To incorporate this effect into the gap
equation, we modify the Eq.\BF\ by introduction of some weights
$\rho _\kappa $, corresponding to the gauge bosons ${\tilde H}_\kappa
$, which become small for the heavy gauge bosons,
$$
M={f^2\over4\pi ^2}\sum_\kappa \rho _\kappa {\tilde T}_\kappa
      \left[1+{M^2\over\Lambda ^2}\ln{M^2\over\Lambda ^2}\right]M {\tilde
T}_\kappa .
\eqn\BG
$$
Here, we take $\Lambda $ for the mass of the lightest gauge bosons.
Note that the dominant part of Eq.\BG, which comes from the lightest
gauge bosons,
is not modified, \ie, $\rho _{\rm \ lightest}=1$, and for the other
gauge bosons, $\rho _\kappa <1$. Eq.\BF\ is the case that all
$\rho _\kappa =1$.

Now, we consider quark mass matrices of up- and down- sectors.
These two mass matrices satisfy the same Eq.\BG,
since the horizontal interactions are common to both sectors.
We can see from Eq.\BG\ that the mass differences among the
generations depend upon the breaking pattern of
the horizontal symmetry, $\rho _\kappa $ and ${\tilde T}_\kappa $.
Eq.\BG\ cannot, however, discriminate between up-sector and
down-sector, which leads us to the same mass matrices.

This result, however, can be avoided by taking into account that the
vertical gauge forces influence each sector in different way.
In fact, the vertical gauge interaction $U(1)_Y$ can discriminate
between up- and down-sectors, giving small
corrections to the gap equations dominated by 4-Fermi or horizontal
gauge interactions.
Many people\refmark{\BKS,\KSY,\Yama} evaluate the effective coupling $G_{\rm
eff}$,
$$
{G_{\rm eff} \over G_{\rm cr}} = {G \over G_{\rm cr}}+
{3 \over 8\pi ^2} g_1^2(\Lambda)Y_{\rm L}Y_{\rm R},
\eqn\BH
$$
where $g_1(\Lambda)$ is $U(1)_Y$ running coupling constant at $\Lambda $, say,
$3g_1^2(10{\rm TeV})/8\pi^2$ $\sim 5\times10^{-3}$,
$Y_{\rm L(R)}$ is hypercharge of left(right) handed quarks, and
$$
G = {f^2 \over 4\pi ^2}.
\eqn\BI
$$
$G_{\rm cr}$ is a critical coupling constant for the dynamical mass
generation of Eq.\BG.
One of the present authors evaluated\refmark{\NT} similar expression
in terms of horizontal gauge coupling $f$ by calculating two loop
diagrams with QED corrections.

At a glance, this small correction could not induce a large mass
splitting between up- and down-sectors, especially between top and bottom.
At this point, however, it is quite important to note that
our model is to be a near critical system, that is,
the horizontal gauge coupling constant is taken to be very close to
the critical one. This type of fine-tuning is in general needed to
relate a high energy scale theory to a low energy physics.
In fact, if the coupling constant is not fine-tuned in our case, the gap
equation \BG\ has a solution $M\sim O(\Lambda)$.
As long as we expect the mass scale
$\Lambda _{\rm SM}$ of the standard model, a fine-tuning of the order of
$O(\Lambda _{\rm SM}/\Lambda)$ is needed.

It has been pointed out that a small perturbation as
mentioned above may be enhanced under the fine-tuned
system.\refmark{\NNT} To make this point clear, we
consider two systems with different coupling constants $G_U$ and $G_D$,
where $G_U$ is a little larger than $G_D$. One example is a case that
$G_U$$ > G_{cr}$$ > G_D$, which implies that solutions of the gap equations
are $M_U\not=0$ and $M_D=0$.
Another example is a case that $G_U$$ > G_D$$ > G_{cr}$ and
$G_U - G_{cr} \gg  G_D - G_{cr}$.
We rewrite the gap equation Eq.\BG\ as follows,
$$
{1 \over G_{cr}} - {1 \over G} \sim  -{M^2 \over \Lambda ^2}\ln{M^2 \over
\Lambda ^2},
\eqn\BJ
$$
where we neglect matrix form for simplicity. Eq.\BJ\
indicates that $M_U\gg M_D$ in this example.

In this paper, we adopt the later case, where $G_{U(D)}$
is a effective coupling for up(down)-sectors.
Indeed, the $\uy$ interaction is
attractive for up-sector and repulsive for down-sector.
$M_U \gg  M_D$ can
be realized with a fine-tuning of $G$.
To obtain difference between $m_t$ and $m_b$ by $\uy$, we must
take $\Lambda \sim 20$TeV, which implies that $G_U - G_{cr}\sim
O(10^{-3})$ and $m_t\sim 150$TeV.\refmark{\BKS}

\chapter{Linearizing Approximation and Eigenvalue Problem}
In this section, we shall investigate the relationship between breaking
patterns of horizontal symmetry and mass matrices.
In the preceding section, we have shown that
solutions of Eq.\BG\ can provide large difference between $M_U$ and
$M_D$. Noting that $M^2/\Lambda ^2\ll 1$, Eq.\BG\ is
satisfied mainly by cancellation between linear terms of $M$ on each
hand side. This indicates that the matrix form of $M$ is mainly
determined by linear parts of Eq.\BG, though
the scale of $M$ is determined by its nonlinear terms, as seen
from Eq.\BJ. Therefore, we neglect nonlinear terms of the gap equation for
a while in order to study forms of mass matrices.

The linearized gap equation is
$$
M=G\sum_\kappa \rho _\kappa {\tilde T}_\kappa M{\tilde T}_\kappa .
\eqn\CA
$$
Note that this approximation is exact on the critical points.
The essential point is that to solve Eq.\CA\ is nothing but an eigenvalue
problem,
in which the coupling constant $G$ and mass matrix $M$ correspond to an
eigenvalue and an eigenvector respectively.
At a glance, only discrete and finite number of couplings would be allowed,
because we consider now only linear parts of gap equation.

When Eq.\CA\ has several eigenvalues, which should we select?
This is a problem how to search for the most stable solutions.
The answer is given by choosing the smallest eigenvalue for a near
critical system.
Because the fine-tuned solution is less stable than the others, it
must be one and only nontrivial solution. It means that the eigenvalue
of the fine-tuned solution is the smallest.
In fact, supposing that there are two positive eigenvalues
$G_1$ and $G_2$ ($G_1<G_2$), solutions of the full gap equation
\BG\ with coupling constant $G$ in the following five cases are conceivable:
\item{\rm a)} $G<G_1<G_2$ \nextline
There are no nontrivial solutions.
\item{\rm b)} $G_1\lsim G<G_2$ \nextline
There is one fine-tuned solution, corresponding to $G_1$ eigenmode.
\item{\rm c)}$G_1<G<G_2$ \nextline
There is one solution, corresponding to $G_1$ eigenmode.
\item{\rm d)}$G_1<G_2\lsim G$ \nextline
There is one fine-tuned solution corresponding to $G_2$ eigenmode
beside another solution corresponding to $G_1$.
\item{\rm e)}$G_1<G_2<G$ \nextline
There are two solutions.

\noindent
By noting that the fine-tuned state is less stable than the others,
$G_1$ mode in case d) turns out to be more stable than fine-tuned
$G_2$ mode and chosen. Then, b) is the only case that we want.

We apply the above rule to simple examples, such as $\rho _\kappa =\{0, 1\}$
and $\Tt_\kappa ={1\over2} \lambda _\kappa $, where $\lambda _\kappa $
is Gellmann matrix.
These examples mean that the horizontal gauge bosons $\Ht_\kappa $
corresponding to $\rho _\kappa =0$ have very large masses
and $\Ht_\kappa $ corresponding to $\rho _\kappa =1$ have small masses
$\Lambda $. Moreover, some symmetries are assumed to be
survived at $\Lambda $, for example, $\uc$, $\ub \times \ua$ and $\ua$.
\item{\rm i)}$\uc$ case \nextline
In this case, all $\rho _\kappa =1$. The Eq.\CA\ has one positive eigenvalue,
$$
G={3\over4}, \qquad\qquad\qquad M\sim \pmatrix{1&&\cr&1&\cr&&1\cr}.
\eqn\CB
$$
This result is natural since there is a global $\uc$ horizontal
 symmetry. It is, however, undesirable phenomenologically.
\item{\rm ii)}$\ub\times \ua$ case \nextline
Here, we take $\rho _{1,2,3,8}=1$ and $\rho _{4,5,6,7}=0$. The Eq.\CA\ has
following two positive eigenvalues,
$$
G={6\over5}, \qquad\qquad\qquad M\sim \pmatrix{1&&\cr&1&\cr&&0\cr},
\eqn\CC
$$
$$
G=3, \qquad\qquad\qquad M\sim \pmatrix{0&&\cr&0&\cr&&1\cr}.
\eqn\CD
$$
The solution \CD\ is desirable phenomenologically,
which means only one generation is massive.
However, it is ruled out
by the principle that the smallest eigenvalue must be selected. Then, we
have the phenomenologically undesirable solution \CC\ in this case.
\item{\rm iii)}$\ua$ case \nextline
We take $\rho _8=1$ and the others are zero. The Eq.\CA\ has
following two positive eigenvalues,
$$
G=3, \qquad\qquad\qquad M\sim \pmatrix{0&&\cr&0&\cr&&1\cr},
\eqn\CE
$$
$$
G=12, \qquad\qquad\qquad M\sim \pmatrix{a&b&\cr c&d&\cr&&0\cr},
\eqn\CF
$$
where $a,b,c$ and $d$ are arbitrary parameters. The smallest eigenvalue
solution \CE\ is phenomenologically desirable.
The realization of the solution \CE\ is also understandable if we note that
$\lambda _8\sim {\rm diag}(1,1,-2)$, which implies that third
generation feels the $H_8$ interaction twice than the others do.

\chapter{Origin of Difference between Up and Down Sectors}
{}From now on, we shall study the origin of mixings. In the preceding
section,
 we have shown that the linearized gap equation \CA\ well describes matrix
form of $M$. Starting from this linearizing approximation, we
take account of the effects of nonlinear terms in the gap equation
Eq.\BG. For simplicity, we apply the linearizing approximation to
down-sector as in the preceding section, because $G_D$ is closer to
$G_{cr}$ than $G_U$. In this sense, we deal with only up-sector below.

The nonlinear terms of \BG\ play two important roles in our model:
One is to determine scale of mass matrix $M$, as mentioned above.
The other is to generate mixing, which means that a solution $M$ of
Eq.\BG\ is not proportional to the solution of linearized Eq.\CA.
In order to understand this intuitively, we introduce an iteration method
for solving the gap equation \BG.

At first, it is assumed that the linearized gap equation \CA\ has
eigenvalues $G_i$ ($G_1<G_2<\cdots<G_n$) and corresponding eigenvectors
$M_i$. We rewrite the gap equations \CA\ and \BG\ as
$$
M = GA_0[M],
\eqn\DA
$$
and
$$
M = G\left(A_0[M]+A_1[M]\right),
\eqn\DB
$$
respectively, where $A_0$ is a linear operation and $A_1$ is a nonlinear
one. One operation of $A_0$ on $M_i$ is
$$
M_i \longrightarrow GA_0[M_i] = {G \over G_i}M_i,
\eqn\DC
$$
which means that, if $G>G_i$, the corresponding mode $M_i$ grows with
iteration, and if $G<G_i$, it dumps. The mode corresponding to the
smallest eigenvalue $G_1$ is most dominant, as mentioned in the preceding
section, since it has the largest factor $G/G_1$ in \DC. If $G\not=G_i$,
$M_i$ diverges or vanishes by repetition of \DA. This indicates that
Eq.\DA\ demands that $G$ is one of the eigenvalues and $M$ is the
corresponding eigenvector.

When $G$ does not belong to eigenvalues, \DB\ must be considered.
In neglecting matrix form in \BG, nonlinear term $A_1(M)$ in
\DB\ has negative contribution. If $G<G_i$, $M_i$ dumps faster than \DA\
and the solution is $M=0$. If $G>G_i$, the effects of $A_1$ weaken the
degree of
divergence of $M$ in iteration, and the scale of the solutions
is determined when effects of $A_1$ and $A_0$ cancels each other.
Larger $G/G_i$ requires larger scale of $M$ since the effect of $A_1$
should be sufficiently large for canceling that of $A_0$.

$A_1$ has another effect in general, which rotates eigenmodes and
generates mixings. Starting from $M_1$ with $G>G_1$, which is the most
dominant mode, one operation of \DB\ leads us to
$$
G\left(A_0[M_1]+A_1[M_1]\right)\longrightarrow
 \left({G\over G_1}+\delta \right)M_1+({\rm other\ modes}),
\eqn\DD
$$
where $\delta $ is $O(M_1^2/\Lambda ^2)$. It is important to point out that
other
modes are smaller than $M_1$ with many iterations because $A_1[M]$ is
$O(M^2/\Lambda ^2)$ and $G/G_i$ is smaller than $G/G_1$. The scale of other
modes is determined by the balance of effects between $A_0$ and $A_1$.
In general, when $G/G_i$ is larger or $A_1$ generates $M_i$ mode more,
$M_i$ mode is larger.

Experiments show non-proportionality between $M_U$ and $M_D$,
for example,
$m_c/m_t:m_s/m_b \sim 1:3$. Can we realize such
non-proportionality? The answer is that, if $G_1$ and $G_2$ are
sufficiently close, it is possible. As shown above, when $G/G_1$ and
$G/G_2$ are not so different, the suppression of $M_2$ mode weaken and this
mode survives at last.

\chapter{Down-quark Diagonalizing Base}
The gap equations for up- and down-sectors are
$$
	\eqalign{
M_D&=G_D\sum_\kappa \rho _\kappa {\tilde T}_\kappa M_D{\tilde T}_\kappa ,\cr
M_U&=G_U\sum_\kappa \rho _\kappa {\tilde T}_\kappa \left[1+{M_U^2\over\Lambda
^2}\ln{M_U^2\over\Lambda ^2}\right]M_U
{\tilde T}_\kappa .\cr
	}\eqn\FA
$$
$\rho _\kappa $ was denoted in our previous paper\refmark{\NNT} as
$$
\rho _\kappa =\ln{\Lambda ^2\over M_\kappa ^2},
	\eqn\FB
$$
with cut-off $\Lambda $ and horizontal gauge boson masses $M_\kappa $.

Before entering into detailed discussion, we will briefly summarize the base
transformations of the gap equations.
Eq.\FA\ contain multi-index quantities $({\tilde T}_\kappa)_{ij}$ and $M_{ij}$
where $i,j$ denote the quark generations running $1\sim 3$ and $\kappa $ the
$SU(3)$ generators running $1\sim 8$.
Corresponding to these two types of indices, we deal with two different types
of bases: (a) the quark base ($i,j,\dots$) and (b) the HG boson base
($\kappa ,\kappa ',\dots;\alpha ,\beta ,\dots$).
 \item{\rm(a)}The quark base is transformed by $SU(3)$ matrix.
Quark mass matrices can always be diagonalized by this base transformation.
 \item{\rm(b)}The HG boson base is transformed by $O(8)$ matrix.
We have already used this type of rotation to diagonalize the HG boson mass
matrix and $T_\kappa \equiv {1\over2}\lambda _\kappa $ was replaced by
${\tilde T}_\kappa $.

Type (a) transformations $U$ constitute a proper subset (subgroup) of the type
(b) transformations.
Indeed, type (b) transformations have greater degrees of freedom than (a).
We can always rewrite any $SU(3)$ matrix $U$ as its adjoint representation
$R_{\alpha \beta }(\in O(8))$ defined by
$$
R_{\alpha \beta }={1\over2}\Tr U^{\dagger}\lambda ^\alpha U\lambda ^\beta .
	\eqn\FE
$$
By virtue of the orthogonality of the Gellmann matrices
$\Tr\lambda ^\alpha \lambda ^\beta =2\delta ^{\alpha \beta }$, Eq.\FE\
can also be written as follows;
$$
R_{\alpha \beta }\lambda ^\beta =U^{\dagger}\lambda ^\alpha U.
	\eqn\FF
$$

However, we cannot express every $O(8)$ rotations in the form of
Eq.\FE. Especially, horizontal mixing angles, \ie, a rotation matrix
$O_{\kappa \alpha }$ , which transforms the HG bosons from the
standard Gellmann base into the mass diagonal base, can not always be
compensated for by the quark base transformation.
In other words, ${\tilde T}_\kappa \equiv{1\over2}O_{\kappa \alpha
}\lambda ^\alpha $ never be written in the form of the right-hand side
of Eq.\FF\ in general.
We write down again the gap equations \FA, indicating the horizontal mixing
angels manifestly;
$$\eqalignno{
M_D&=x_\kappa O_{\kappa \alpha }O_{\kappa \beta }\lambda ^\alpha
M_D\lambda ^\beta ,&\eqname{\FG}\cr
M_U&=\xi x_\kappa O_{\kappa \alpha }O_{\kappa \beta }\lambda ^\alpha
\left[1+{M_U^2\over\Lambda ^2}\ln{M_U^2\over\Lambda ^2}\right]M_U
\lambda ^\beta ,
&\eqname{\FH}\cr
}$$
where $x_\kappa ={1\over4}G_D\rho _\kappa $ and $\xi=G_U/G_D$.
$\xi$ is evaluated from Eq.\BH,
$$
\xi=1+{\rm O}(10^{-3}).
\eqn\FD
$$

Now, let us define down-quark diagonalizing base (DDB), which is selected to
diagonalize the down-sector quark mass matrix.
Suppose that we successfully solve the Eq.\FG\ for given $x_\kappa $,
$O_{\kappa \alpha }$ and obtain a solution $M_D$.
There exists a unitary transformation $U_D$ which
diagonalize $M_D$. Transforming Eq.\FG\ and \FH\ by $U_D$, we obtain
$$\eqalignno{
D_D&=x_\kappa {\tilde O}_{\kappa \alpha }{\tilde O}_{\kappa \beta }
\lambda ^\alpha D_D\lambda ^\beta ,&
\eqname{\FI}\cr
M_U&=\xi x_\kappa {\tilde O}_{\kappa \alpha }{\tilde O}_{\kappa \beta }\lambda
^\alpha
\left[1+{M_U^2\over\Lambda ^2}\ln{M_U^2\over\Lambda ^2}\right]M_U\lambda ^\beta
,&
\eqname{\FJ}\cr
}$$
where we denote ${\tilde O}=OR$, using a rotation matrix $R$ defined by
$$
R_{\alpha \beta }={1\over2}\Tr U_D^\dagger\lambda ^\alpha U_D\lambda ^\beta ,
\eqn\FK
$$
and redefine the quark mass matrices
$$\eqalignno{
U_D^\dagger M_DU_D&\longrightarrow D_D=
\pmatrix{m_d&&\cr&m_s&\cr&&m_b\cr},&
\eqname{\FL}\cr
U_D^\dagger M_UU_D&\longrightarrow M_U=
V_{KM}^\dagger\pmatrix{m_u&&\cr&m_c&\cr&&m_t\cr}V_{KM}.&
\eqname{\FM}\cr
}$$
Here, $V_{KM}$ is CKM matrix. Note that, in this
expression, all ambiguous unphysical degrees of freedom are fixed,\ie,
the mass matrices are written only by the quark masses and mixing angles.

\chapter{Down-sector Equation}
In the preceding section, we obtained the set of equations \FI\ and
\FJ\ in DDB, which the quark mass parameters, \ie, the masses and the mixings
should satisfy.
Our task is now to find out weight parameters $x_\kappa $
and rotation matrix ${\tilde O}_{\kappa \alpha }$, which give rise to
phenomenologically acceptable quark mass parameters.
Note that $x_\kappa $ and ${\tilde O}_{\kappa \alpha }$
are constrained by the DDB condition.
In fact, the vanishness of the off-diagonal elements of $D_D$ in Eq.\FI\
imposes six real conditions on 8 + 28 degrees of freedom of $x_\kappa $ and
${\tilde O}_{\kappa \alpha }$.
Moreover, taking account of the experimental values of the diagonal
elements (down quark masses), two additional real conditions exist.
Note that Eq.\FI\ does not determine an overall scale of solutions.

Since the down-sector equation \FI\ is a linear equation for $D_D$, we can
rewrite it in the form of 9-dimensional eigenvalue problem to see the above
conditions in more detail.
Let us define nine orthonormal matrix units $\sigma _p$ of the hermitian
matrices:

$$\eqalign{
\sigma _1=\pmatrix{1&&\cr&0&\cr&&0},\qquad
\sigma _4={1\over\sqrt2}\pmatrix{0&0&0\cr0&0&1\cr0&1&0},\qquad
\sigma _7={1\over\sqrt2}\pmatrix{0&0&0\cr0&0&-i\cr0&i&0},\cr
\sigma _2=\pmatrix{0&&\cr&1&\cr&&0},\qquad
\sigma _5={1\over\sqrt2}\pmatrix{0&0&1\cr0&0&0\cr1&0&0},\qquad
\sigma _8={1\over\sqrt2}\pmatrix{0&0&-i\cr0&0&0\cr i&0&0},\cr
\sigma _3=\pmatrix{0&&\cr&0&\cr&&1},\qquad
\sigma _6={1\over\sqrt2}\pmatrix{0&1&0\cr1&0&0\cr0&0&0},\qquad
\sigma _9={1\over\sqrt2}\pmatrix{0&-i&0\cr i&0&0\cr0&0&0},\cr
}\eqn\GA
$$
satisfying $\Tr\sigma _p\sigma _q=\delta _{pq}$.
The down-sector mass matrix can be represented in terms of $\sigma _p$,
$$
D_D=d_p\sigma _p,\qquad\quad d_p=\Tr\sigma _pD_D,
\eqn\GB
$$
where $d_p$'s are components corresponding to $\sigma _p$'s.
Taking trace of Eq.\FI\ multiplied with $\sigma _p$, we obtain the eigen
equation for 9-dimensional vector $d_p$:
$$
d_p=A_{pq}d_q,
\eqn\GC
$$
where $A_{pq}$ defined as
$$
A_{pq}=x_\kappa {\tilde O}_{\kappa \alpha }{\tilde O}_{\kappa \beta }
\Tr\sigma _p\lambda _\alpha \sigma _q\lambda _\beta .
\eqn\GD
$$
The 9$\times $9 matrix $A_{pq}$, which is real symmetric and traceless
by definition, contains all physical information of the mass matrix of
the horizontal gauge boson corresponding to $x_\kappa $ and ${\tilde
O}_{\kappa \alpha }$.

Generally, $A_{pq}$ has several eigenvalues.
We introduce eigenvalue $\eta $ explicitly into Eq.\GC;
$$
d_p=\eta A_{pq}d_q.
\eqn\GE
$$
By construction of $A_{pq}$, Eq.\GE\ must have the solution of
$d_p=(m_d,m_s,m_b,0,0,0,0,0,0)$ with $\eta =1$, which corresponds to
the down quark diagonalized solution (DDS) of \FL.
In addition to this, $\eta \not=1$ solutions can also be realized for gauge
coupling $\eta G_D$. In order for DDS to be chosen, we require that
the DDS \FL\ should be the most stable solution, which corresponds to
the smallest eigenvalue and is realized for the weakest gauge coupling.
If there exist solutions for $\eta <1$, they will dominate as was seen
in sect.4.
Therefore, Eq.\GE\ should not have $\eta <1$ solutions.

Consequently, we can summarize the following DDB conditions for the matrix
$A_{pq}$.
\item{\rm (I)}$A_{pq}$ should have an eigenvector
$d_p=(m_d,m_s,m_b,0,0,0,0,0,0)$ with eigenvalue 1.
\item{\rm (II)}The eigen equation \GE\ has only $\eta \geq 1$
eigenvalues or negative.

\chapter{A Toy Model}
Let us apply our formulation to some simple cases.
We assume that only three HG bosons are light, which have only non-zero
$x_\kappa$ of $x_3,\ x_6,\ x_8$. We also constraint
${\tilde O}_{\kappa \alpha}$ to be $3\times 3$ matrix with $\kappa$ and
$\alpha$ being 3, 6, 8. The corresponding three Gellmann matrices are;
$$
\lambda _3=              \pmatrix{ 1&  &  \cr   &-1&  \cr   &  & 0},\quad
\lambda _6=              \pmatrix{ 0& 0& 0\cr  0& 0& 1\cr  0& 1& 0},\quad
\lambda _8={1\over\sqrt3}\pmatrix{ 1&  &  \cr   & 1&  \cr   &  & 2}.
\eqn\HA
$$
Under this assumption, the first generation does not couple with the other
generations.
In addition to the diagonal elements (2,2) and (3,3), only real part
of (2,3) element in Eq.\FI\ gives nontrivial constraint:
\def\OO#1#2{{\tilde O}_{\kappa #1}{\tilde O}_{\kappa #2}}
$$\eqalign{
m_s &=x_\kappa \left[\left(\OO33+{1\over3}\OO88-{2\over\sqrt3}\OO38\right)m_s
+\OO66m_b\right],\cr
m_b &=x_\kappa \left[\OO66m_s+{4\over3}\OO88m_b\right],\cr
0  &=x_\kappa \left[\left({1\over\sqrt3}\OO68-\OO36\right)m_s
-{2\over\sqrt3}\OO68m_b\right].\cr
}\eqn\HB
$$
Since $m_d$ is negligible compared with $m_s$ and $m_b$, and the first
generation is decoupled from the other two generations, we take
$m_d=0$ by hand.
It means that the constraint from (1,1) element in Eq.\FI\ is trivial.

By setting $m_s/m_b$ to its experimental value, we can calculate
$x_\kappa $ for given ${\tilde O}_{\kappa \alpha }$ using the above
equations \HB. We parametrize
${\tilde O}_{\kappa \alpha }$ by Euler-like angles:
\def\sa{\sin\theta_{36}}
\def\ca{\cos\theta_{36}}
\def\sb{\sin\theta_{38}}
\def\cb{\cos\theta_{38}}
\def\sc{\sin\theta_{68}}
\def\cc{\cos\theta_{68}}
$$
\tilde O = \pmatrix{\ca&-\sa&0\cr\sa&\ca&0\cr 0&0&1}
           \pmatrix{\cb&0&-\sb\cr 0&1&0\cr\sb&0&\cb\cr}
           \pmatrix{1&0&0\cr 0&\cc&-\sc\cr 0&\sc&\cc\cr}.
\eqn\HC
$$
\FIG\figB{ A $\theta _{68}-\theta _{38}$ plane with $\theta_{36}=\dg{10}$. }
\FIG\figC{ A magnification of Fig.\figB. }
We search all parameter space of these Euler angles and have found
some allowed region (Fig.\figB,\figC) which satisfies the DDB
conditions (I) and (II), by the following procedures.

\item{1.}Give the Euler angles, and calculate $x_\kappa $ from Eq.\HB.
(For some peculiar values of the Euler angles, Eq.\HB\ is degenerate
so that we cannot obtain $x_\kappa $ from this equation. These values
are presented by 'v' in the figures.)
\item{2.}Check all $x_\kappa $ to be positive.
(The negative $x_\kappa $ is unphysical.
It is indicated as '-' in the figures.)
\item{3.}Construct $A_{pq}$ in Eq.\GD\ for this $x_\kappa $ with the
${\tilde O}_{\kappa \alpha }$, and find its eigenvalues $\eta $ (See Eq.\GE).
If there exists some $\eta $ which is less than 1, DDS is less stable
and not realized (indicated as 'x').
Only the case with $\eta $ being not less than 1 is allowed by the DDB
conditions. There exists at least one eigenvalue which equals to 1.
It is corresponding to DDS and indicated as 'A'.

\noindent
For example, Fig.\figB\ shows a $\theta _{68}-\theta _{38}$ plane with
$\theta _{36}=\dg{10}$, and an allowed region is magnified in Fig.\figC.

We can now solve Eq.\FJ\ for up-sector with allowed Euler angles
obtained above by iteration.
We simply replace matrix ${(1/\Lambda ^2)}\ln{[M_U^2/\Lambda ^2]}$ by
$\zeta ={(1/\Lambda ^2)}\ln{[(M_U)_{33}^2/\Lambda ^2]}$,
since this modification does not affect our result so much;
$$
M_U=\xi x_\kappa {\tilde O}_{\kappa \alpha }{\tilde O}_{\kappa\beta}
\lambda ^\alpha\left[M_U+\zeta M_U^3\right]\lambda ^\beta .
\eqn\HD
$$
Here we present an interesting solution $M_U$, which is not
proportional to $M_D$, as expected, such as
$$
{M_U\over\Lambda }=
\pmatrix{0&0&0\cr0&75.15\times10^{-6}&-1.75\times10^{-6}\cr
0&-1.75\times10^{-6}&0.00876880},
\eqn\HE
$$
with $(\theta _{36}, \theta _{38}, \theta _{68})=(\dg{10}, \dg{-85.765},
\dg{10})$ and
$\xi =1.001$. It means that $m_t=175$GeV, $m_c=1.5$GeV, $m_u=0$ when $\Lambda
=20$TeV.
The above Euler angles correspond to the weights
$\rho_3=0.75008512$, $\rho_6=0.00080017$ and $\rho_8=0.72880273$.
The eigenvalues and eigenvectors of corresponding $A_{pq}$ is as
follows.
\tenpoint
\def\p{\phantom{+}}
$$
\matrix{
     \eta^{(1)}&\eta^{(2)}&\eta^{(3)}&\eta^{(4)}&\eta^{(5)}&\eta^{(6)}\cr
 1.00000000& \p1.01981096& \p1.02345204&  -2.10257169&  -2.01046516&
-1.98033194\cr
&&&&&\cr
     d^{(1)}&d^{(2)}&d^{(3)}&d^{(4)}&d^{(5)}&d^{(6)}\cr
 0.00000000& \p1.00000000& \p0.00000000& \p0.00000000& \p0.00000000&
\p0.00000000\cr
 0.03996804& \p0.00000000& \p0.99915918& \p0.00000000&  -0.00913700&
\p0.00000000\cr
 0.99920096& \p0.00000000&  -0.03996637& \p0.00000000& \p0.00036548&
\p0.00000000\cr
 0.00000000& \p0.00000000& \p0.00914431& \p0.00000000& \p0.99995819&
\p0.00000000\cr
 0.00000000& \p0.00000000& \p0.00000000&  -0.32998315& \p0.00000000&
\p0.94398682\cr
 0.00000000& \p0.00000000& \p0.00000000& \p0.94398682& \p0.00000000&
\p0.32998315\cr
}
$$
\twelvepoint
\centerline{\bf Table.1}
Here, we arrange Table.1 in order of inverse of eigenvalues.
Since $A_{pq}$ is non-vanishing only for $p,q\leq 6$, we regard $A$ as
6$\times $6 matrix.
Eigenvector $d^{(1)}$ with $\eta=1$ corresponds to the down-sector solution.

Let us investigate features of the above solutions in brief.
As shown in sect.4, up-sector solution $M_U$ is formed by mixing a little
$d^{(3)}$ with $d^{(1)}$. These two eigenvalues are close compared
with others except for $\eta ^{(2)}$,
the eigenvector of which is decoupled in the gap equation \FJ.

The present solution can generate CKM mixings. From Eq.\FM, CKM matrix
is given as
$$
V_{KM}=\pmatrix{1&0&0\cr0&1-{1\over2}\alpha^2&-\alpha\cr
0&\alpha&1-{1\over2}\alpha}\quad {\rm with}\quad \alpha=0.0002,
\eqn\HF
$$
off-diagonal elements of which arise from mixing $d^{(3)}$ eigenvector.
Since we have not searched all allowed regions, we could not here realize
agreement between Eq.\HF\ and the experimental results.

\chapter{Conclusions and Discussions}
We have discussed the dynamical mass generation and the possibility of
CKM mixings by the broken horizontal gauge interactions.
The essential point of generating the difference between $M_U$ and
$M_D$ is that the fine-tuned system is perturbed by the radiative
$U(1)_Y$ corrections. The above results are induced by the mechanism
that nonlinear terms of the gap equation mix eigenvectors of the
linearized gap equation. We should emphasize that the above results are
caused though the horizontal interactions themselves do not
discriminate between up- and down-sectors.

In the above model, there arise no Cabbibo angle and CP violating phase.
In order to get realistic CKM matrix, we should consider the first
generation and horizontal interactions corresponding to
$\lambda_\kappa$ with imaginary elements.

In sect.3, we adopted the linearized gap equation for down sector.
This is justified by confirming that the solution of nonlinear equation
goes to that of linearized equation in the limit of $G\rightarrow G_{cr}$.
We explain it by using the model in sect.7 with $\theta _{36}=\dg{10}$,
$\theta _{38}=\dg{-85.75}$ and $\theta _{68}=\dg{10}$.
We define the eigenvectors in this case as $d^{(i)}$ like in sect.7.
The solution of linearized equation is $d^{(1)}$ and that of
nonlinear equation is a linear combination of $d^{(1)}$ and $d^{(3)}$
approximately.
Table.2 shows the ratio $r$ of $d^{(1)}$ to $d^{(3)}$ for the
solutions of nonlinear equation with several coupling $G$.

$$
\matrix{
G&1.0004&1.0006&1.0008&1.0010&1.0012\cr
r&-0.0110&-0.0166&-0.0223&-0.0281&-0.0340\cr
}
$$
\centerline{\bf Table.2}

Finally, we discuss on the horizontal breaking scale $\Lambda$.
In the model of sect.7, $m_t=175$GeV demands that $\Lambda=20$TeV.
It does not agree with the present experiments of FCNC, which requires
that $\Lambda>1000$TeV.
However, the diagonal horizontal interactions are free from this constraint.
Off-diagonal interactions do not satisfy this constraint in general.
Fortunately, in the above solution, small weight $\rho_6$ obtained above
suppress FCNC due to $H_6$ interaction to some extent.

\acknowledge{
The present authors wish to thank S.Tanaka for valuable discussions
and a careful reading of the manuscript.}

\refout

\figout

\vskip 5cm

\epsfxsize=8cm
\centerline{\epsfbox{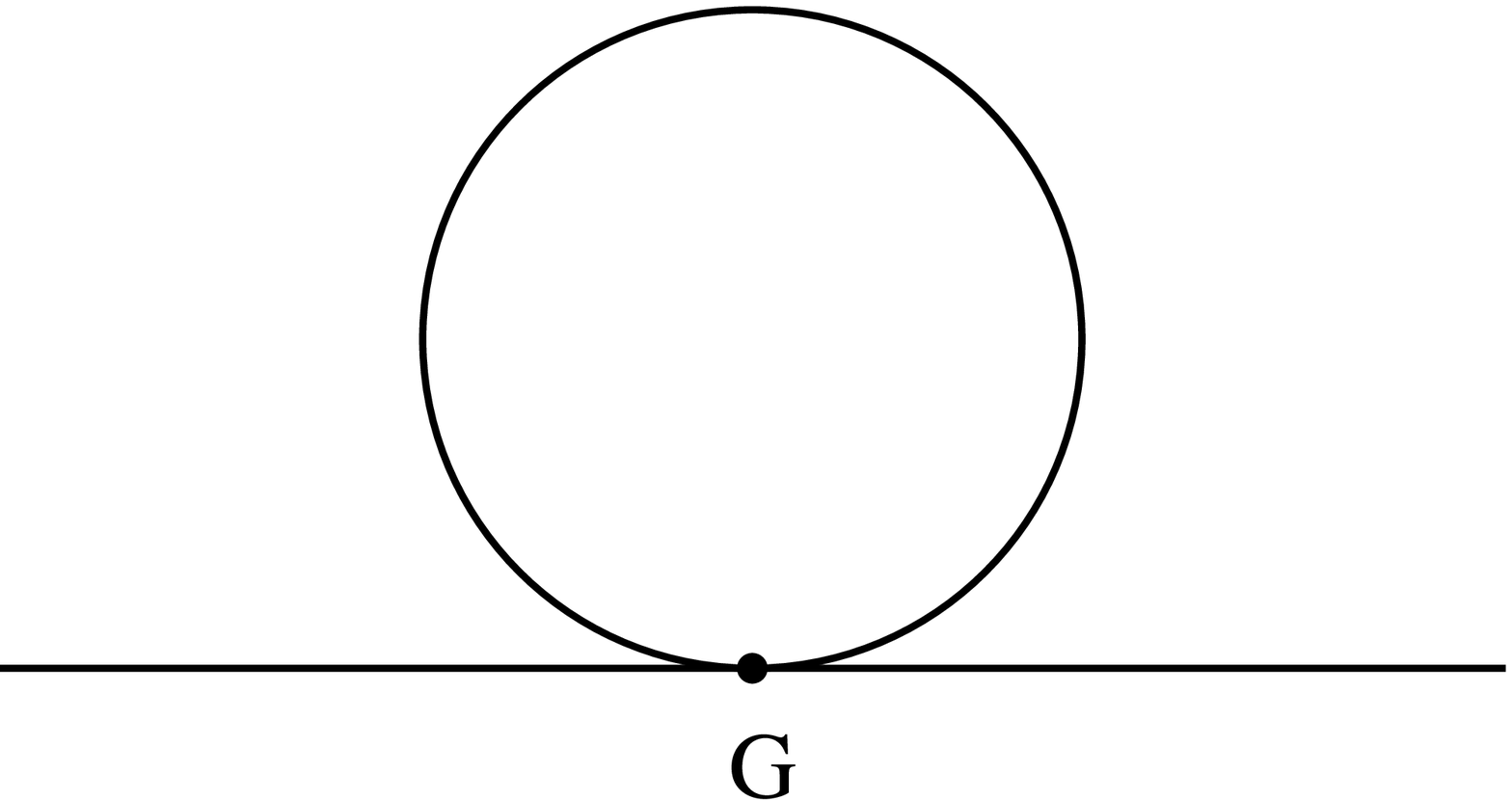}}
\nextline\nextline
\centerline{Fig.\figA}

\endpage

\def\uncatcodespecials{\def\do##1{\catcode`##1=12 }\dospecials}

\def\verbatim{\begingroup\tt\uncatcodespecials
\obeyspaces\doverbatim}
\newcount\balance
{\catcode`<=1 \catcode`>=2 \catcode`\{=12 \catcode`\}=12
\gdef\doverbatim{<\balance=1\verbatimloop>
\gdef\verbatimloop#1<\def\next<#1\verbatimloop>%
\if#1{\advance\balance by1
\else\if#1}\advance\balance by-1
\ifnum\balance=0\let\next=\endgroup\fi\fi\fi\next>>

\baselineskip=10pt

$\theta_{36}=\dg{10}$\qquad
$\theta_{38}=\dg{0}\pm\dg{90}$\qquad
$\theta_{68}=\dg{0}\pm\dg{90}$\nextline\nextline\nextline
\noindent
\hbox{~~~~~~~~}\verbatim{ -90   -60   -30    0    +30   +60
+90}\quad$\theta_{68}$
\nextline
\hbox{~~~~~~~~~~~}\verbatim{+     +     +     +     +     +     +}\nextline
\verbatim{-90  + }\verbatim{--------------------v----------------}\nextline
\hbox{~~~~~~~~~~~}\verbatim{--------------------A----------------}\nextline
\hbox{~~~~~~~~~~~}\verbatim{--------------------A----------------}\nextline
\verbatim{-60  + }\verbatim{--------------------A----------------}\nextline
\hbox{~~~~~~~~~~~}\verbatim{--------------------x----------------}\nextline
\hbox{~~~~~~~~~~~}\verbatim{---------------------x---------------}\nextline
\verbatim{-30  + }\verbatim{-A-----------------------------------}\nextline
\hbox{~~~~~~~~~~~}\verbatim{------------------A------------------}\nextline
\hbox{~~~~~~~~~~~}\verbatim{------------------A------------------}\nextline
\verbatim{0    + }\verbatim{v-----------------A-----------------v}\nextline
\hbox{~~~~~~~~~~~}\verbatim{------------------A------------------}\nextline
\hbox{~~~~~~~~~~~}\verbatim{------------------A------------------}\nextline
\verbatim{+30  + }\verbatim{------------------A----------------A-}\nextline
\hbox{~~~~~~~~~~~}\verbatim{------------------x------------------}\nextline
\hbox{~~~~~~~~~~~}\verbatim{-----------------xx------------------}\nextline
\verbatim{+60  + }\verbatim{-------------------------------------}\nextline
\hbox{~~~~~~~~~~~}\verbatim{----------------A--------------------}\nextline
\hbox{~~~~~~~~~~~}\verbatim{----------------A-----------------A--}\nextline
\verbatim{+90  + }\verbatim{----------------v--------------------}\nextline
\nextline
\hbox{~}$\theta_{38}$\nextline
\centerline{Fig.\figB}

\nextline
\nextline

$\theta_{36}=\dg{10}$\qquad
$\theta_{38}=\dg{-60}\pm\dg{40}$\qquad
$\theta_{68}=\dg{13}\pm\dg{5.7}$\nextline\nextline\nextline
\noindent
\hbox{~~~~~~~~~~~}\verbatim{        10.0      13.0      16.0
}\quad$\theta_{68}$
\nextline
\hbox{~~~~~~~~~~~}\verbatim{    +    +    +    +    +    +    +    }\nextline
\verbatim{-100 + }\verbatim{---------A-----------------------------}\nextline
\hbox{~~~~~~~~~~~}\verbatim{---------A-----------------------------}\nextline
\hbox{~~~~~~~~~~~}\verbatim{---------A-----------------------------}\nextline
\hbox{~~~~~~~~~~~}\verbatim{---------------------------------------}\nextline
\hbox{~~~~~~~~~~~}\verbatim{---------A-----------------------------}\nextline
\verbatim{-80  + }\verbatim{---------A-----------------------------}\nextline
\hbox{~~~~~~~~~~~}\verbatim{---------AA----------------------------}\nextline
\hbox{~~~~~~~~~~~}\verbatim{--------AAA----------------------------}\nextline
\hbox{~~~~~~~~~~~}\verbatim{--------AAAA---------------------------}\nextline
\hbox{~~~~~~~~~~~}\verbatim{-------AAAAAAA-------------------------}\nextline
\verbatim{-60  + }\verbatim{-------AAAAAAAA------------------------}\nextline
\hbox{~~~~~~~~~~~}\verbatim{------xxxxxxxxxxx----------------------}\nextline
\hbox{~~~~~~~~~~~}\verbatim{------xxxxxxxxxxxxx--------------------}\nextline
\hbox{~~~~~~~~~~~}\verbatim{------xxxxxxxxxxxxxxxx-----------------}\nextline
\hbox{~~~~~~~~~~~}\verbatim{--------xxxxxxxxxxxxxxxxx--------------}\nextline
\verbatim{-40  + }\verbatim{----------xxxxxxxxxxxxxxxxxx-----------}\nextline
\hbox{~~~~~~~~~~~}\verbatim{-----------------xxxxxxxxxxxxxxxx------}\nextline
\hbox{~~~~~~~~~~~}\verbatim{----------------------------xxxxxxxxxx-}\nextline
\hbox{~~~~~~~~~~~}\verbatim{---------------------------------------}\nextline
\hbox{~~~~~~~~~~~}\verbatim{---------------------------------------}\nextline
\verbatim{-20  + }\verbatim{---------------------------------------}\nextline
\nextline
\hbox{~}$\theta_{38}$\nextline
\centerline{Fig.\figC}

\bye